\let\oldvec\vec
\documentclass[runningheads,citeauthoryear,oribibl]{llncs}
\let\vec\oldvec

\usepackage[T1]{fontenc}
\usepackage{graphicx}
\usepackage{amssymb}
\usepackage{amsmath}
\usepackage{algorithm}
\usepackage[noend]{algpseudocode}
\usepackage[group-separator={,}]{siunitx}
\usepackage{multirow}
\usepackage{complexity}
\usepackage{xspace}
\usepackage{multicol}
\usepackage{cutwin}

\newtheorem{mytheorem}{Theorem}

\newtheorem{myproposition}{Proposition}
\newtheorem{mydefinition}{Definition}
\newtheorem{mycorollary}{Corollary}
\newtheorem{myexample}{Example}

\newcommand{\myqed}{\mbox{$\square$}}

\newcolumntype{C}{>{\centering\arraybackslash}p{2cm}}

\begin{document}

\title{Two Algorithms for Additive and \\ Fair Division of Mixed Manna}
\titlerunning{Two Algorithms for Additive and Fair Division of Mixed Manna}

\author{Martin Aleksandrov \and Toby Walsh}
\authorrunning{M. Aleksandrov, T. Walsh}

\institute{Technical University Berlin, Germany \\
\email{\{martin.aleksandrov,toby.walsh\}@tu-berlin.de}}

\maketitle

\begin{abstract}
We consider a fair division model in which agents have positive, zero and negative utilities for items. For this model, we analyse one existing fairness property - EFX - and three new and related properties - EFX$_0$, EFX$^3$ and EF1$^3$ - in combination with Pareto-optimality. With general utilities, we give a modified version of an existing algorithm for computing an EF1$^3$ allocation. With $-\alpha/0/\alpha$ utilities, this algorithm returns an EFX$^3$ and PO allocation. With absolute identical utilities, we give a new algorithm for an EFX and PO allocation. With $-\alpha/0/\beta$ utilities, this algorithm also returns such an allocation. We report some new impossibility results as well. 
\end{abstract} 

\section{Introduction}\label{sec:intro}

Fair division of indivisible items lies on the intersection of fields such as social choice, computer science and algorithmic economics \cite{chevaleyre2006}. Though a large body of work is devoted to the case when the items are goods (e.g.\ \cite{brams1996,steinhaus1948,moulin2003,young1995}), there is a rapidly growing interest in the case of mixed manna (e.g.\ \cite{aziz2019popropone,caragiannis2012,sandomirskiy2019minimal}). In a mixed manna, each item can be classified as \emph{mixed} (i.e.\ some agents strictly like it and other agents strictly dislike it), \emph{good} (i.e.\ all agents weakly like it and some agents strictly like it), \emph{bad} (i.e.\ all agents weakly dislike it and some agents strictly dislike it) or \emph{dummy} (i.e.\ all agents are indifferent to it). 

An active line of fair division research currently focuses on approximations of envy-freeness (i.e.\ no agent envies another one) \cite{foley1967}. For example, Aziz et al.\ \cite{aziz2019gc} proposed two such approximations for mixed manna: EF1 and EFX. EF1 requires that an agent's envy for another agent's bundle is eliminated by removing some item from these agents' bundles. EFX strengthens EF1 to any non-zero valued item in these bundles, increasing the agent's utility or decreasing the other agent's utility. However, they study only EF1 and identify improving our understanding of EFX as an important open problem for mixed manna:

\begin{quote}
{\em ``Our work paves the way for detailed examination of allocation of goods/ chores, and opens up an interesting line of research, with many problems left open to explore. In particular, there are further fairness concepts that could be studied from both existence and complexity issues, most notably envy-freeness up to the least valued item (EFX) \cite{caragiannis2016}.''}
\end{quote}

We make in this paper a step forward in this direction. In particular, we study not only EFX but also {\em new} properties, all stronger than EF1. For example, one such property is {\em envy-freeness by parts up to some item}: EF1$^3$. This ensures EF1 independently for the set of all items, the set of goods and the set of bads (i.e.\ the different parts). Another such property is {\em envy-freeness by parts up to any item}: EFX$^3$. This requires EFX for each of the different parts of the set of items. Yet a third such property is EFX$_0$. This one extends the existing envy-freeness up to any (possibly zero valued) good from \cite{plaut2018} to any (possibly zero valued) bad by relaxing the non-zero marginal requirements in the definition of EFX. We will shortly observe the following relations between these properties.

\begin{center}
EFX$_0$\hspace{0.15cm}$\Rightarrow$\hspace{0.15cm}EFX\hspace{0.5cm}
EFX$^3$\hspace{0.15cm}$\Rightarrow$\hspace{0.15cm}EFX\hspace{0.5cm} EF1$^3$\hspace{0.15cm}$\Rightarrow$\hspace{0.15cm}EF1 \hspace{0.5cm} EFX$^3$\hspace{0.15cm}$\Rightarrow$\hspace{0.15cm}EF1$^3$
\end{center}

We analyse these properties in isolation and also in combination with an efficiency criterion such as Pareto-optimality (PO). PO ensures that we cannot make an agent happier without making another one unhappier. More precisely, we ask in our work whether combinations of these properties can be guaranteed, and also how to do this when it is possible. Our analysis covers three common domains for \emph{additive} (i.e.\ an agent's utility for a set of items is a sum of their utilities for the items in the set) utility functions: \emph{general} (i.e.\ each utility is real-valued), \emph{absolute identical} (i.e.\ for each item, the agents' utilities have identical magnitudes but may have different signs) as well as \emph{ternary} (i.e.\ each utility is $-\alpha$, $0$ or $\beta$ for some $\alpha,\beta\in\mathbb{R}_{>0}$). 

Each of these domains can be observed in practice.  For instance, if a machine can perform a certain task faster than some pre-specified amount of time, then its utility for the task is positive and, otherwise, it is negative. Thus, multiple machines can have mixed utilities for tasks. Further, consider a market where items have prices and agents sell or buy items. In this context, the agent's utilities for an item have identical magnitudes but different signs. Finally, a special case of ternary utilities is when each agent have utility $-1$, $0$, or $1$ for every item. This is practical because we need simply to elicit whether agents like, dislike or are indifferent to each item. A real-world setting with such utilities is the food bank problem studied in \cite{aleksandrov2015ijcai}. 

We give some related work, formal preliminaries and motivation in Sections~\ref{sec:work},~\ref{sec:pre} and~\ref{sec:mot}, respectively. In Section~\ref{sec:gen}, we give a polynomial-time algorithm (i.e.\ Algorithm~\ref{alg:mdrr}) for computing an EF1$^3$ allocation with general utilities. We also prove that an EFX$^3$ allocation, or an EFX$_0$ allocation might not exist even with ternary identical utilities. In Section~\ref{sec:ident}, we give a polynomial-time algorithm (i.e.\ Algorithm~\ref{alg:minimax}) for computing an EFX and PO allocation with absolute identical utilities, and show that Algorithm~\ref{alg:mdrr} returns an EF1$^3$ and PO allocation. In Section~\ref{sec:ter}, we show that Algorithm~\ref{alg:mdrr} returns an EF1$^3$ and PO allocation with ternary utilities, whereas Algorithm~\ref{alg:minimax} returns an EFX and PO allocation. Finally, we give a summary in Section~\ref{sec:con}.

\section{Related work}\label{sec:work}

For indivisible goods, EF1 was defined by Budish \cite{budish2011}. Caragiannis et al. \cite{caragiannis2016} proposed EFX. It remains an open question of whether EFX allocations exist in problems with general utilities. Recently, Amanatidis et al. \cite{amanatidis2020maximum} proved that EFX allocations exist in \emph{2-value} (i.e.\ each utility takes one of two values) problems. In contrast, we show that EFX and PO allocations exist in problems with ternary (i.e.\ $-\alpha/0/\beta$) utilities, which are special cases of 3-value problems. Barman, Murthy and Vaish \cite{barman2018fe} presented a pseudo-polynomial time algorithm for EF1 and PO allocations. Barman et al. \cite{barman2018} gave an algorithm for EFX and PO allocations in problems with identical utilities. Plaut and Roughgarden \cite{plaut2018} proved that the {\em leximin} solution from \cite{dubins1961} is also EFX and PO in this domain. Although this solution maximizes the minimum agent's utility (i.e.\ the egalitarian welfare), it is intractable to find in general \cite{dobzinski2013}. In our work, we give a polynomial-time algorithm for EFX and PO allocations in problems with absolute identical utilities, and show that this welfare and EFX$^3$ are incompatible. 

For mixed manna, Aziz et al.\ \cite{aziz2019gc} proposed EF1 and EFX. They gave the double round-robin algorithm that returns EF1 allocations. Unfortunately, these are not guaranteed to satisfy PO. They also gave a polynomial-time algorithm that returns allocations which are EF1 and PO in the case of \num{2} agents. Aziz and Rey \cite{aziz2019group} gave a ``ternary flow'' algorithm for leximin, EFX and PO allocations with $-\alpha/0/\alpha$ utilities. With $-\alpha/0/\beta$ utilities, we discuss that these might sadly violate EFX$^3$ even when $\alpha=1,\beta=1$, or EFX when $\alpha=2,\beta=1$. By comparison, we give a modified version of the double round-robin algorithm that returns EF1$^3$ allocations in problems with general utilities, EF1$^3$ and PO allocations in problems with absolute identical utilities and EFX$^3$ and PO allocations in problems with $-\alpha/0/\alpha$ utilities. Other works of divisible manna are \cite{bogomolnaia2019div,bogomolnaia2016gab,bogomolnaia2017comp}, and approximations of envy-freeness for indivisible goods are \cite{amanatidis2018,caragiannis2016,lipton2004}. In contrast, we study some new approximations and the case of indivisible manna. 

\section{Formal preliminaries}\label{sec:pre}

We consider a set $[n]=\lbrace 1,\ldots, n\rbrace$ of $n\in\mathbb{N}_{\geq 2}$ agents and a set $[m]=\lbrace 1,\ldots,m\rbrace$ of $m\in\mathbb{N}_{\geq 1}$ indivisible items. We assume that each agent $a\in [n]$ have some \emph{utility} function $u_a:2^{[m]}\rightarrow\mathbb{R}$. Thus, they assign some utility $u_a(M)$ to each bundle $M\subseteq [m]$. We write $u_a(o)$ for $u_a(\lbrace o\rbrace)$. We say that $u_a$ is \emph{additive} if, for each $M\subseteq [m]$, $u_a(M)=\sum_{o\in M} u_a(o)$. We also write $u(M)$ if, for each other agent $b\in [n]$, $u_a(M)=u_b(M)$. 

With additive utility functions, the set of items $[m]$ can be partitioned into \emph{mixed items}, \emph{goods}, \emph{bads} and \emph{dummies}. Respectively, we write $[m]^{\pm}=\lbrace o\in [m]|\exists a\in [n]: u_a(o)>0,\exists b\in [n]:u_b(o)<0\rbrace$, $[m]^+=\lbrace o\in [m]|\forall a\in [n]: u_a(o)\geq 0,\exists b\in [n]:u_b(o)>0\rbrace$, $[m]^-=\lbrace o\in [m]|\forall a\in [n]:u_a(o)\leq 0,\exists b\in [n]:u_b(o)<0\rbrace$ and $[m]^0=\lbrace o\in [m]|\forall a\in [n]:u_a(o)=0\rbrace$ for the sets of these items. We refer to an item $o$ from $[m]^+$ as a \emph{pure good} if $\forall a\in [n]: u_a(o)>0$. Also, we refer to an item $o$ from $[m]^-$ as a \emph{pure bad} if $\forall a\in [n]: u_a(o)<0$.

We say that agents have \emph{general} additive utilities if, for each $a\in [n]$ and each $o\in [m]$, $u_a(o)$ could be any number from $\mathbb{R}$. Further, we say that they have \emph{absolute identical} additive utilities if, for each $o\in [m]$, $|u_a(o)|=|u_b(o)|$ where $a,b\in [n]$, or \emph{identical} additive utilities if, for each $o\in [m]$, $u_a(o)=u_b(o)$ where $a,b\in [n]$. Finally, we say that agents have \emph{ternary} additive utilities if, for each $a\in [n]$ and each $o\in [m]$, $u_a(o)\in \lbrace -\alpha,0,\beta\rbrace$ for some $\alpha,\beta\in\mathbb{R}_{>0}$. 

An \emph{(complete) allocation} $A=(A_1,\ldots,A_n)$ is such that (1) $A_a$ is the set of items allocated to agent $a\in [n]$, (2) $\bigcup_{a\in [n]} A_a=[m]$ and (3) $A_a\cap A_b=\emptyset$ for each $a,b\in[n]$ with $a\neq b$. We consider several properties for allocations.

\paragraph{\bf Envy-freeness up to one item}\label{par:ef}

Envy-freeness up to one item requires that an agent's envy for another's bundle is eliminated by removing an item from the bundles of these agents. Two notions for our model that are based on this idea are EF1 and EFX \cite{aziz2019gc}.

\begin{mydefinition} $(${\em EF1}$)$
An allocation $A$ is \emph{envy-free up to some item} if, for each $a,b\in [n]$, $u_a(A_a)\geq u_a(A_b)$ or $\exists o\in A_a\cup A_b$ such that $u_a(A_a\setminus\lbrace o\rbrace)\geq u_a(A_b\setminus\lbrace o\rbrace)$.
\end{mydefinition} 

\begin{mydefinition}$(${\em EFX}$)$
An allocation $A$ is \emph{envy-free up to any non-zero valued item} if, for each $a, b\in [n]$, (1) $\forall o\in A_a$ such that $u_a(A_a)$ $<$ $u_a(A_a\setminus\lbrace o\rbrace)$: $u_a(A_a\setminus\lbrace o\rbrace)\geq u_a(A_b)$ and (2) $\forall o\in A_b$ such that $u_a(A_b)>u_a(A_b\setminus\lbrace o\rbrace)$: $u_a(A_a)\geq u_a(A_b\setminus\lbrace o\rbrace)$.
\end{mydefinition} 

Plaut and Roughgarden \cite{plaut2018} considered a variant of EFX for goods where, for any given pair of agents, the removed item may be valued with zero utility by the envy agent. Kyropoulou et al.\ \cite{kyropoulou2019} referred to this one as EFX$_0$. We adapt this property to our model. 

\begin{mydefinition}$(${\em EFX$_0$}$)$
An allocation $A$ is \emph{envy-free up to any item} if, for each $a, b\in [n]$, (1) $\forall o\in A_a$ such that $u_a(A_a)$ $\leq$ $u_a(A_a\setminus\lbrace o\rbrace)$: $u_a(A_a\setminus\lbrace o\rbrace)\geq u_a(A_b)$ and (2) $\forall o\in A_b$ such that $u_a(A_b)\geq u_a(A_b\setminus\lbrace o\rbrace)$: $u_a(A_a)\geq u_a(A_b\setminus\lbrace o\rbrace)$.
\end{mydefinition} 

An allocation that is EFX$_0$ further satisfies EFX. Also, EFX is stronger than EF1. It is well-known that the opposite relations might not hold. 

\paragraph{\bf Envy-freeness by parts}\label{par:exfthree}

Let $A=(A_1,\ldots,A_n)$ be a given allocation. We let $A^+_a=\lbrace o\in A_a|u_a(o)>0\rbrace$ and $A^-_a=\lbrace o\in A_a|u_a(o)<0\rbrace$ for each $a\in [n]$. Envy-freeness by parts up to one item ensures that EF1 (or EFX) is satisfied in each of the allocations $A$, $A^+=(A^+_1,\ldots,A^+_n)$ and $A^-=(A^-_1,\ldots,A^-_n)$.

\begin{mydefinition}$(${\em EF1$^3$}$)$
An allocation $A$ is \emph{envy-free by parts up to some item} $($\emph{EF1-EF1-EF1 or EF1$^3$}$)$ if the following conditions hold: (1) $A$ is EF1, (2) $A^+$ is EF1 and (3) $A^-$ is EF1.
\end{mydefinition} 

\begin{mydefinition}$(${\em EFX$^3$}$)$
An allocation $A$ is \emph{envy-free by parts up to any item} $($\emph{EFX-EFX-EFX or EFX$^3$}$)$ if the following conditions hold: (1) $A$ is EFX, (2) $A^+$ is EFX and (3) $A^-$ is EFX.
\end{mydefinition} 

With just goods (bads), EF1$^3$ (EFX$^3$) is EF1 (EFX). With mixed manna, an allocation that is EF1$^3$ also satisfies EF1, one that is EFX$^3$ satisfies EFX, and one that is EFX$^3$ satisfies EF1$^3$. The reverse implications might not be true. 

\paragraph{\bf Pareto-optimality}\label{par:po}

We also study each of these fairness properties in combination with an efficiency criterion such as Pareto-optimality (PO), proposed a long time ago by Vilfredo Pareto \cite{pareto1896}. 

\begin{mydefinition}$(${\em PO}$)$
An allocation $A$ is \emph{Pareto-optimal} if there is no allocation $B$ that \emph{Pareto-improves} $A$, i.e.\ $\forall a\in [n]$: $u_a(B_a)\geq u_a(A_a)$ and $\exists b\in [n]$: $u_b(B_b)> u_b(A_b)$.
\end{mydefinition} 

\section{Further motivation}\label{sec:mot}

\opencutright
\renewcommand\windowpagestuff{%
\hspace{0.075\columnwidth}
\includegraphics[height=4.25cm,width=0.85\columnwidth]{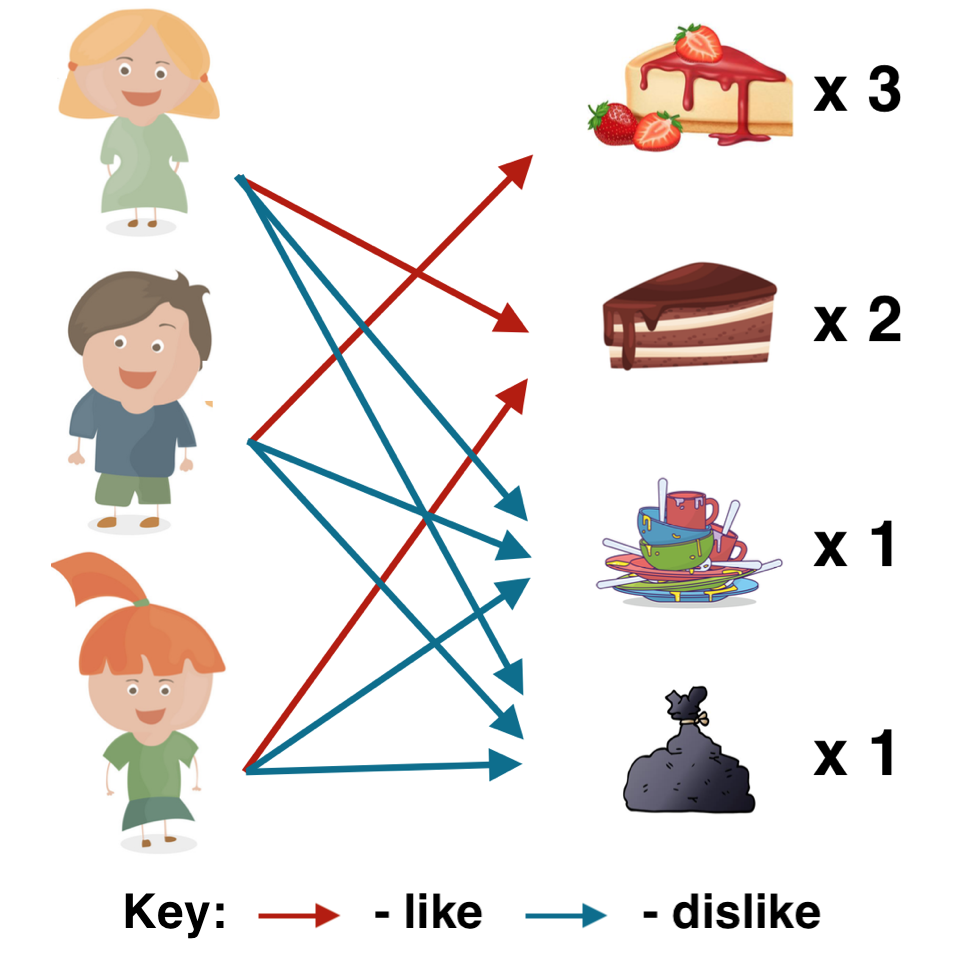}
}
\begin{cutout}{9}{0.5\columnwidth}{0pt}{10}
We next further motivate the new properties EF1$^3$ and EFX$^3$ by means of a simple example. Consider a birthday party where Bob invites his new friends Alice and Mary. Bob has \num{3} pieces of his favourite \emph{strawberry cake} (value is $1$) and \num{2} pieces of the less favorable to him \emph{chocolate cake} (value is $0$). Bob also hopes that some of his guests would be willing to help him \emph{washing up the dishes} and \emph{throwing away the garbage} after the party.  Alice and Mary arrive and it turns out that both like only chocolate cake (value is $1$), and dislike any of the household chores (value is $-1$) as does Bob. How shall we allocate the \num{5} goods (i.e.\ all pieces of cake) and the \num{2} chores? \newline\indent For EF1 (EFX) and PO, we shall give the strawberry cake to Bob and one piece of the chocolate cake to each of Alice and Mary. As a result, Bob gets utility $3$ whereas Alice and Mary get each utility $1$. If we want to maximize the egalitarian welfare, we should assign both chores to Bob. Doing so preserves EF1 (EFX) and PO for all items. However, it violates EF1$^3$ (EFX$^3$). Indeed, Bob might be unhappy simply because they have to do both chores instead of sharing them with Alice and Mary. This means that an EF1 (EFX) allocation might not satisfy EF1$^3$ (EFX$^3$). In contrast, achieving EF1$^3$ (EFX$^3$) avoids assigning both chores to Bob. For example, asking Bob to wash up the dishes and Alice to throw away the garbage, or vice versa is EF1$^3$ (EFX$^3$). Other such options share the chores between Bob and Mary, and Alice and Mary. However, none of these maximizes the egalitarian welfare. This means that EF1$^1$ (EFX$^3$) is incompatible with this objective.
\end{cutout}

\section{General additive utilities}\label{sec:gen}

We begin with general utilities. An EF1 allocation in this domain can be computed in $O(\max\lbrace m^2,mn\rbrace)$ time. For this purpose, we can use the existing \emph{double round-robin} algorithm from \cite{aziz2019gc}. However, this algorithm may fail to guarantee PO because an agent might pick a bad for which some other agent have zero utility. 

\begin{myexample}\label{exp:gen}
Consider \num{2} agents and \num{2} items, say $a$ and $b$. Define the utilities as follows: $u_1(a)=-1$, $u_1(b)=-1$ and $u_2(a)=-1$, $u_2(b)=0$. In this problem, the double round-robin algorithm is simply a round-robin rule with some strict priority ordering of the agents. Wlog, let agent 1 pick before agent 2. Wlog, let agent 1 pick $b$. Now, agent 2 can only pick $a$. The returned allocation gives utility $-1$ to agent 1 and utility $-1$ to agent 2. By swapping these items, agent 1 receive utility $-1$ and agent 2 receive utility $0$. Clearly, this is a Pareto-improvement.
\myqed
\end{myexample}

In response, we modify slightly the double round-robin algorithm by adding an extra preliminary phase where each dummy item/non-pure bad is allocated to an agent who has zero utility for it: Algorithm~\ref{alg:mdrr}. As we show, this modified version gives us an EF1$^3$ allocation that is PO not only with $-1/0/1$ utilities but also with any ternary utilities, as well as with absolute identical utilities. 

\begin{mytheorem}\label{thm:efonethreegen}
With general utilities, Algorithm~\ref{alg:mdrr} returns an EF1$^3$ allocation in $O(\max\lbrace m^2,mn\rbrace)$ time. 
\end{mytheorem}

\begin{algorithm}
\caption{An EF1$^3$ allocation (see the Appendix for a complete version).}\label{alg:mdrr}
\begin{algorithmic}[1]
\Procedure{Modified Double Round-Robin}{$[n],[m],(u_1,\ldots,u_n)$} 
\State $M^0\gets\lbrace o\in [m]|\forall b\in [n]:u_b(t)\leq 0,\exists c\in [n]:u_c(t)=0\rbrace$
\State $\forall a\in [n]: A_a\gets\emptyset$
\For{$t\in M^0$} \Comment{allocate all dummies/non-pure bads}
\State pick $a\in\lbrace b\in [n]|u_b(t)=0\rbrace$
\State $A_a\gets A_a\cup\lbrace t\rbrace$
\EndFor
\State $B\gets ${\sc Double Round-Robin}($[n],[m]\setminus M^0,(u_1,\ldots,u_n)$)
\State \Return $(A_1\cup B_1,\ldots,A_n\cup B_n)$ 
\EndProcedure
\end{algorithmic}
\end{algorithm}

\begin{myproof}
The double round-robin algorithm returns an EF1 allocation, and so $B$ is EF1. Consider $B^+$ and $B^-$. Let there be $qn-p$ pure bads for some $q,p\in\mathbb{N}$ with $p<n$. The algorithm creates $p$ ``fake'' dummy items for which each agent has utility $0$, and adds them to the set of pure bads. Hence, the number of items in this set becomes $qn$. Thus, the agents come in a round-robin fashion according to some ordering of the agents, say $(1,\ldots,n-1,n)$, and pick their most preferred item in this set (i.e.\ all pure bads and ``fake'' dummies) until all of them are allocated. This is EF1 for the pure bads. Hence, $B^-$ is EF1.

Further, the agents come in a round-robin fashion by following the reversed ordering, i.e.\ $(n,n-1,\ldots,1)$, and pick their most preferred good until all mixed items and goods are allocated. If an agent has no available item which gives them strictly positive utility, they pretend to pick a new ``fake'' dummy item for which they have utility $0$. This is EF1 for the mixed items and goods. Hence, $B^+$ is also EF1 which implies that $B$ is EF1$^3$. Finally, extending $B$ to all items, by allocating each dummy item/non-pure bad to someone who holds zero utility, preserves EF1$^3$. This means that the returned allocation is EF1$^3$.
\myqed
\end{myproof}

We move to stronger properties. For example, EFX$^3$ allocations in our setting might not exist. The rationale behind this is that an agent may get their least valued bad in an attempt of achieving EFX for the bads. As a result, removing this bad from their bundle might not be sufficient to eliminate their envy of some other agent who receive positive utility for a good and a bad. 

\begin{myproposition}\label{pro:impefxthree}
There are problems with \num{2} agents and ternary identical utilities for \num{1} pure goods and \num{2} pure bads, in which \emph{no} allocation is EFX$^3$.
\end{myproposition}  

\begin{myproof} 
Suppose that there are \num{2} agents and \num{3} items. We define the utilities as follows: $u(a)=-1$, $u(b)=-1$, $u(c)=2$. We note that one EFX allocation gives items $a$, $b$ and $c$ to agent 1 and no items to agent 2. However, there is no allocation that satisfies EFX$^3$.

We observe that there are two EFX allocations of the pure bads, i.e.\ $A=(\lbrace a\rbrace,\lbrace b\rbrace)$ and $B=(\lbrace b\rbrace,\lbrace a\rbrace)$. Further, we observe that there are two EFX allocations of the pure good, i.e.\ $C=(\lbrace c\rbrace,\emptyset)$ and $D=(\emptyset,\lbrace c\rbrace)$. By the symmetry of the utilities, we consider only $A$, $C$ and $D$.

If we unite (``agent-wise'') $A$ and $C$, then $u(A_2\cup C_2\setminus\lbrace b\rbrace)=0<1=u(A_1\cup C_1)$. Therefore, the union of $A$ and $C$ is not EFX and, therefore, EFX$^3$. If we unite $A$ and $D$, then $u(A_1\cup D_1\setminus\lbrace a\rbrace)=0<1=u(A_2\cup D_2)$. Again, the union of $A$ and $D$ violates EFX$^3$. Similarly, for $B$, $C$ and $D$.
\myqed
\end{myproof}

By comparison, EFX allocations exist in 2-value problems with goods \cite{amanatidis2020maximum}. It follows immediately that EFX$^3$ allocations exist in such problems. From this perspective, we feel that our impossibility result compares favorably to this possibility result because such allocations may not exist in 2-value problems with goods and bads.

Even more, this result also implies that no EFX allocation satisfies EF1$^3$ and no EF1$^3$ allocation satisfies EFX in some problems with identical and ternary utilities. As a consequence, any allocation that could be returned by Algorithm~\ref{alg:mdrr} might violate EFX. These implications are also true for the stronger version EFX$_0$ in problems where such allocations exist.

However, EFX$_0$ allocations might also not always exist. The reason for this might be the presence of dummies. One may argue that such items could be removed. However, some web-applications on Spliddit for example ask agents to announce items (e.g.\ inherited items) and utilities but the system has no access to the actual items and, therefore, cannot remove the dummies \cite{caragiannis2016}.

\begin{myproposition}\label{pro:impefxzero}
There are problems with \num{2} agents and ternary identical utilities for \num{1} pure good and \num{1} dummy, in which \emph{no} allocation is EFX$_0$.
\end{myproposition}  

\begin{myproof} 
Suppose that there are \num{2} agents and \num{2} items, say $a$ and $b$. We define the utilities as follows: $u(a)=1$ and $u(b)=0$. We argue that there is no EFX$_0$ allocation in this problem. To see this, we make two observations. Firstly, with the given set of items, it is impossible that both agents obtain the same utility, as the individual utilities are integers and their sum is odd. Secondly, EFX$_0$ for the agents in this problem where a dummy item is present requires that both agents have the same utility. This follows by the definition of EFX$_0$. \myqed
\end{myproof}

This result is perhaps interesting because EFX$_0$ allocations exist in problems with \num{2} agents and general utilities for goods \cite{plaut2018}, or \emph{any} number of agents and $0/1$ utilities \cite{amanatidis2020maximum}. By Propositions~\ref{pro:impefxthree} and~\ref{pro:impefxzero}, it follows that neither EFX$^3$ nor EFX$_0$ can be achieved in combination with PO, or even a weaker efficiency notion such as \emph{completeness} (i.e.\ all items are allocated), in general. 

\section{Absolute identical additive utilities}\label{sec:ident}

We continue with absolute identical utilities. Requiring such utilities is not as strong as requiring just identical utilities. To see this, consider agents 1, 2 and items $a$, $b$. Define the utilities as $u_1(a)=3$, $u_1(b)=2$ and $u_2(a)=3$, $u_2(b)=-2$. The absolute values of these utilities are identical but their cardinal values are not, e.g.\ $|u_1(b)|=|u_2(b)|=2$ but $u_1(b)=2, u_2(b)=-2$.

By Proposition~\ref{pro:impefxthree}, EF1$^3$ and EFX are incompatible in this domain. Nevertheless, we can combine each of them with PO. For example, Algorithm~\ref{alg:mdrr} returns an allocation that satisfies PO besides EF1$^3$. The key reason for this result is that in such problems there are no items that are bads (goods) for some agents and dummy for other agents. 

\begin{mytheorem}\label{thm:efonethreepoid}
With absolute identical utilities, Algorithm~\ref{alg:mdrr} returns an EF1$^3$ and PO allocation. 
\end{mytheorem}

\begin{myproof}
EF1$^3$ follows by Theorem~\ref{thm:efonethreegen}. We note that each allocation that gives at least one mixed item to an agent who values it strictly negatively can be Pareto-improved by moving this item to an agent who values it strictly positively. Therefore, such an allocation is not Pareto-optimal. We also note that each other allocation, including the returned one, maximizes the sum of agents' utilities because it achieves the maximum utility for each individual item. Such an allocation is always Pareto-optimal.
\myqed
\end{myproof}

At the same time, we can compute an EFX and PO allocation in polynomial time. For this task, we propose a \emph{new} algorithm: Algorithm~\ref{alg:minimax}. We let $M(o)=\max_{a\in [n]} u_a(o)$ denote the maximum utility that an agent derives from item $o$. Further, let us arrange the items in non-increasing absolute maximum utility order by using the following tie-breaking rule.

{\em Ordering} $\sigma_m$: Wlog, $|M(1)|\geq \ldots\geq |M(m)|$. Initialize $\sigma_m$ to $(1,\ldots,m)$. While there are two items $s$ and $t$ from $[m]$ such that $|M(s)|=|M(t)|$, $M(s)>0$, $M(t)<0$ and $t$ is right before $s$ in $\sigma_m$, do move $s$ right before $t$ in $\sigma_m$. Thus, within items with the same maximum absolute utility, $\sigma_m$ gives higher priority to the mixed items/goods than to the pure bads.

Algorithm~\ref{alg:minimax} allocates the items one-by-one in such an ordering $\sigma_m$. If the current item $t$ is mixed or pure good, then Algorithm~\ref{alg:minimax} gives it to an agent who has currently the minimum utility among the agents who like the item. If item $t$ is pure bad, then Algorithm~\ref{alg:minimax} gives it to an agent who has currently the maximum utility. Otherwise, it gives item $t$ to an agent with zero utility.

\begin{mytheorem}\label{thm:efxpoid}
With absolute identical utilities, Algorithm~\ref{alg:minimax} returns an EFX and PO allocation in $O(\max\lbrace m\log m,mn\rbrace)$ time. 
\end{mytheorem}

\begin{algorithm}
\caption{An EFX and PO allocation.}\label{alg:minimax}
\begin{algorithmic}[1]
\Procedure{Minimax}{$[n],[m],(u_1,\ldots,u_n)$} 
\State compute the ordering $\sigma_m$ \Comment{see right above}
\State $\forall a\in [n]: A_a\gets\emptyset$
\For{$t\in \sigma_m$}
\If {$t$ is mixed item or good} 
\State $N\gets \lbrace b\in [n]|u_b(t)>0\rbrace$
\State $\mbox{MinUtil}(A)\gets \lbrace b\in N|u_b(A_b)=\min_{c\in N} u_c(A_c)\rbrace$
\State pick $a\in\mbox{MinUtil}(A)$
\ElsIf {$t$ is pure bad}
\State $\mbox{MaxUtil}(A)\gets \lbrace  b\in [n]|u_b(A_b)=\max_{c\in [n]} u_c(A_c)\rbrace$
\State pick $a\in\mbox{MaxUtil}(A)$
\Else \Comment{$t$ is dummy item or non-pure bad}
\State pick $a\in\lbrace b\in [n]| u_b(t)=0\rbrace$
\EndIf
\State $A_a\gets A_a\cup\lbrace t\rbrace$
\EndFor
\State \Return $(A_1,\ldots,A_n)$ 
\EndProcedure
\end{algorithmic}
\end{algorithm}

\begin{myproof}
For $t\in [m]$, we let $A^t$ denote the partially constructed allocation of items $1$ to $t$. Pareto-optimality of $A^t$ follows by the same arguments as in Theorem~\ref{thm:efonethreepoid}, but now applied to the sub-problem of the first $t$ items. As a result, $A^m$ (i.e.\ the returned allocation) satisfies also PO. 

We next prove that $A^t$ is EFX by induction on $t\in [m]$. This will imply the result for EFX of $A^m$ (i.e.\ the returned allocation). In the base case, let $t$ be $1$. The allocation of item $1$ is trivially EFX. In the hypothesis, let $t>1$ and assume that the allocation $A^{t-1}$ is EFX. In the step case, let us consider round $t$. 

Wlog, let the algorithm give item $t$ to agent 1. That is, $A^t_1=A^{t-1}_1\cup\lbrace t\rbrace$ and $A^t_a=A^{t-1}_a$ for each $a\in [n]\setminus\lbrace 1\rbrace$. It follows immediately by the hypothesis that each pair of different agents from $[n]\setminus\lbrace 1\rbrace$ is EFX of each other in $A^t$. We note that $t$ gives positive, negative or zero utility to agent 1. For this reason, we consider three remaining cases for agent $a\in [n]\setminus\lbrace 1\rbrace$ and agent 1. 

\emph{Case 1}: Let $u_1(t)>0$. In this case, $t$ is mixed item or pure good (good) and $u_1(A^t_1)>u_1(A^{t-1}_1)$ holds. Hence, agent 1 remain EFX of agent $a$ by the hypothesis. For this reason, we next show that agent $a$ is EFX of agent 1. We consider two sub-cases depending on whether agent $a$ belong to $N=\lbrace b\in [n]|u_b(t)>0\rbrace$ or not. We note that $1\in N$ holds because of $u_1(t)>0$.

\emph{Sub-case 1 for $a\rightarrow 1$}: Let $a\not\in N$. Hence, $u_a(t)\leq 0$. As a result, $u_a(A^{t-1}_1)\geq u_a(A^t_1)$ holds. Thus, as $A^{t-1}$ is EFX, we derive that $u_a(A^t_a)=u_a(A^{t-1}_a)\geq u_a(A^{t-1}_1\setminus\lbrace o\rbrace)\geq u_a(A^t_1\setminus\lbrace o\rbrace)$ holds for each $o\in A^{t-1}_1$ with $u_a(o)>0$. We also derive $u_a(A^t_a\setminus\lbrace o\rbrace)=u_a(A^{t-1}_a\setminus\lbrace o\rbrace)\geq u_a(A^{t-1}_1)\geq u_a(A^t_1)$ for each $o\in A^t_a$ with $u_a(o)<0$. Hence, agent $a$ is EFX of agent 1.

\emph{Sub-case 2 for $a\rightarrow 1$}: Let $a\in N$. Hence, $u_a(t)>0$. Moreover, $u_a(A^{t-1}_a)\geq u_1(A^{t-1}_1)$ by the selection rule of the algorithm. For each item $o\in A^{t-1}_1$, we have $u_1(o)=u_a(o)$ if $o$ is pure good, pure bad or dummy item, and $u_1(o)\geq u_a(o)$ if $o$ is mixed item. Therefore, $u_1(A^{t-1}_1)\geq u_a(A^{t-1}_1)$ or agent $a$ is envy-free of agent 1 in $A^{t-1}$. 

We derive $u_a(A^t_a)=u_a(A^{t-1}_a)\geq u_a(A^{t-1}_1)=u_a(A^t_1\setminus\lbrace t\rbrace)$ because $A^t_a=A^{t-1}_a$ and $A^t_1=A^{t-1}_1\cup\lbrace t\rbrace$. Furthermore, $u_a(A^t_1\setminus\lbrace t\rbrace)\geq u_a(A^t_1\setminus\lbrace o\rbrace)$ for each $o\in A^{t-1}_1$ with $u_a(o)>0$ because $u_a(o)\geq u_a(t)$ holds due to the ordering of items used by the algorithm.

We now show EFX of the bads. We have $u_a(A^t_a\setminus\lbrace o\rbrace)=u_a(A^{t-1}_a\setminus\lbrace o\rbrace)\geq u_a(A^{t-1}_1)+u_a(t)=u_a(A^t_1)$ for each $o\in A^{t-1}_a$ with $u_a(o)<0$ because $|u_a(o)|\geq u_a(t)$ holds due to the ordering of items used by the algorithm. Hence, agent $a$ is EFX of agent 1.

\emph{Case 2}: Let $u_1(t)<0$. In this case, $t$ is pure bad and $u_a(A^t_1)<u_a(A^{t-1}_1)$ holds. That is, agent 1's utility decreases. By the hypothesis, it follows that agent $a$ remain EFX of agent 1 in $A^t$. For this reason, we only show that agent 1 remain EFX of agent $a$.

\emph{$1\rightarrow a$}: We have $u_1(A^{t-1}_1)\geq u_a(A^{t-1}_a)$ by the selection rule of the algorithm. For each item $o\in A^{t-1}_a$, we have $u_a(o)=u_1(o)$ if $o$ is pure good, pure bad or dummy item, and $u_a(o)\geq u_1(o)$ if $o$ is mixed item. We conclude $u_a(A^{t-1}_a)\geq u_1(A^{t-1}_a)$ and, therefore, $u_1(A^{t-1}_1)\geq u_1(A^{t-1}_a)$. Hence, agent $1$ is envy-free of agent $a$ in $A^{t-1}$.

Additionally, it follows that $u_1(A^t_1\setminus\lbrace t\rbrace)\geq u_1(A^t_a)$ holds because $A^t_1\setminus\lbrace t\rbrace=A^{t-1}_1$ and $A^t_a=A^{t-1}_a$. Due to the order of the items, we have $|u_1(b)|\geq |u_1(t)|$ for each $b\in A^t_1$ with $u_1(b)<0$. Hence, $u_1(A^t_1\setminus\lbrace b\rbrace)\geq u_1(A^t_1\setminus\lbrace t\rbrace)\geq u_1(A^t_a)$ for each $b\in A^t_1$ with $u_1(b)<0$. 

At the same time, $u_1(A^{t-1}_1)\geq u_1(A^{t-1}_a\setminus\lbrace g\rbrace)$ for each $g\in A^{t-1}_a$ with $u_1(g)>0$. Again, due to the order of the items, $u_1(g)\geq |u_1(t)|$. Therefore, $u_1(A^t_a\setminus\lbrace g\rbrace)\leq u_1(A^t_a)-|u_1(t)|=u_1(A^{t-1}_a)-|u_1(t)|\leq u_1(A^{t-1}_1)-|u_1(t)|=u_1(A^t_1)$. Consequently, $u_1(A^t_1)\geq u_1(A^t_a\setminus\lbrace g\rbrace)$ for each $g\in A^t_a$ with $u_1(g)>0$. 

\emph{Case 3}: Let $u_1(t)=0$. In this case, $t$ is dummy item or non-pure bad. Hence, $u_a(A^{t-1}_1)\geq u_a(A^t_1)$ and $u_1(A^t_1)=u_1(A^{t-1}_1)$ hold. That is, agent 1's utility does not change. By the hypothesis, this means that they remain EFX of each agent $a$ and also each agent $a$ remains EFX of them in $A^t$.

Finally, computing maximum values takes $O(mn)$ time and sorting items takes $O(m\log m)$ time. The loop of the algorithm takes $O(mn)$ time. 
\myqed
\end{myproof}

For problems with identical utilities, Aziz and Ray \cite{aziz2019group} proposed the ``egal-sequential'' algorithm for computing EFX and PO allocations. By Theorem~\ref{thm:efxpoid}, Algorithm~\ref{alg:minimax} also does that. However, we feel that such problems are very restrictive as they do not have mixed items unlike many practical problems.

\begin{mycorollary}\label{cor:identone}
With identical utilities, Algorithm~\ref{alg:minimax} returns an EFX and PO allocation. 
\end{mycorollary}

Algorithm~\ref{alg:minimax} allocates each mixed item/good to an agent who likes it, and each dummy item/non-pure bad to an agent who is indifferent to it. As a consequence, the result in Theorem~\ref{thm:efxpoid} extends to problems where, for each mixed item/good, the likes are identical and, for each pure bad, the dislikes are identical. 

\begin{mycorollary}\label{cor:identtwo}
With identical likes (i.e.\ strictly positive utilities) for each mixed item, identical likes for each good and identical dislikes (i.e.\ strictly negative utilities) for each pure bad, Algorithm~\ref{alg:minimax} returns an EFX and PO allocation.
\end{mycorollary}

\section{Ternary additive utilities}\label{sec:ter}

We end with ternary utilities. That is, each agent's utility for each item is from $\lbrace -\alpha, 0,\beta\rbrace$ where $\alpha,\beta\in\mathbb{R}_{>0}$. We consider two cases for such utilities.

\subsection{Case for any $\alpha,\beta$}

By Proposition~\ref{pro:impefxthree}, it follows that an EFX$^3$ allocation might not exist in some problems even when $\alpha=1$ and $\beta=2$. However, we can compute an EF1$^3$ (notably, also EF1-EFX-EFX) and PO allocation with Algorithm~\ref{alg:mdrr}.

\begin{mytheorem}\label{thm:efonethreepoter}
With ternary utilities from $\lbrace -\alpha,0,\beta \rbrace$ where $\alpha,\beta \in\mathbb{R}_{>0}$, Algorithm~\ref{alg:mdrr} returns an EF1$^3$ and PO allocation. 
\end{mytheorem}

\begin{myproof}
The returned allocation is EF1$^3$ by Theorem~\ref{thm:efonethreegen}. This one achieves the maximum utility for each individual item. Hence, the sum of agents' utilities in it is maximized and equal to $\beta$ multiplied by the number of goods plus $\beta$ multiplied by the number of mixed items minus $\alpha$ multiplied by the number of pure bads. In fact, this holds for each allocation that gives each mixed item/good to an agent who has utility $\beta$, and each dummy/non-pure bad to an agent who has utility $0$. Each other allocation is not PO and does not Pareto-dominate the returned allocation. Hence, the returned one is PO. \myqed
\end{myproof}

One the other hand, we already mentioned after Proposition~\ref{pro:impefxthree} that each allocation returned by Algorithm~\ref{alg:mdrr} in such problems may violate EFX. However, Algorithm~\ref{alg:minimax} returns an EFX and PO allocation in this case.  

\begin{mytheorem}\label{thm:efxpoter}
With ternary utilities from $\lbrace -\alpha,0,\beta \rbrace$ where $\alpha,\beta \in\mathbb{R}_{>0}$, Algorithm~\ref{alg:minimax} returns an EFX and PO allocation. 
\end{mytheorem}

\begin{myproof} This is where the ordering used by the algorithm plays a crucial role. If $\beta\geq\alpha$, we note that all mixed items and goods are allocated before all pure bads and all of these are allocated before the remaining items (i.e.\ dummy items and non-pure bads). If $\beta<\alpha$, we note that all pure bads are allocated before all mixed items and goods and all of these are allocated before the remaining items. Further, we observe that agents have identical likes for each mixed item or each good (i.e.\ $\beta$), and identical dislikes for each pure bad (i.e.\ $-\alpha$). Therefore, the result follows by Corollary~\ref{cor:identtwo}.
\myqed
\end{myproof}

By comparison, the ``ternary flow'' algorithm of Aziz and Rey \cite{aziz2019group} may fail to return an EFX allocation even with $-2/1$ utilities. To see this, simply negate the utilities in the problem from Proposition~\ref{pro:impefxthree}. This algorithm allocates firstly one good to each agent and secondly the bad to one of the agents. This outcome violates EFX. 

\subsection{Case for $\alpha=\beta$}

In this case, we can compute an EFX$^3$ and PO allocation with Algorithm~\ref{alg:mdrr}. Although we consider this a minor result, we find it important because it is the only one in our analysis when EFX$^3$ and PO allocations exist. 

\begin{mytheorem}\label{thm:efxthreepoter}
With ternary utilities from $\lbrace -\alpha,0,\alpha \rbrace$ where $\alpha \in\mathbb{R}_{>0}$, Algorithm~\ref{alg:mdrr} returns an EFX$^3$ and PO allocation. 
\end{mytheorem}

\begin{myproof}
The returned allocation is EF1$^1$ and PO by Theorem~\ref{thm:efonethreepoter}. With general (and, therefore, ternary) utilities, an allocation that is EFX$^3$ also satisfies EF1$^3$ because EFX is a stronger property than EF1, but the opposite implication might not be true. Well, with utilities from $\lbrace -\alpha,0,\alpha \rbrace$, the opposite implication also holds. Indeed, if an allocation is EF1 for a given pair of agents, then removing some good from the envied agent's bundle or removing some bad from the envy agent's bundle eliminates the envy of the envy agent. But, the envy agent likes each such good with $\alpha$ and each such bad with $-\alpha$. Hence, such an allocation is EFX. This implies that an EF1$^3$ allocation is also EFX$^3$ in this domain. \myqed
\end{myproof}

By Theorem~\ref{thm:efxpoter}, Algorithm~\ref{alg:minimax} returns an EFX and PO allocation in this case. However, this one might falsify EFX$^3$ even when $\alpha=1$ (see motivating example). The same holds for the ``ternary flow'' algorithm of Aziz and Rey \cite{aziz2019group} because it maximizes the egalitarian welfare when $\alpha=1$ (see motivating example).

\section{Conclusions}\label{sec:con}

We considered additive and fair division of mixed manna. For this model, we analysed axiomatic properties of allocations such as EFX$_0$, EFX$^3$, EFX, EF1$^3$, EF1 and PO in three utility domains. With general utilities, we showed that an EF1$^3$ allocation exists and gave Algorithm~\ref{alg:mdrr} for computing such an allocation (Theorem~\ref{thm:efonethreegen}). With absolute identical or $-\alpha/0/\beta$ utilities, this algorithm returns an EF1$^3$ and PO allocation (Theorems~\ref{thm:efonethreepoid} and~\ref{thm:efonethreepoter}). With $-\alpha/0/\alpha$ utilities, it returns an EFX$^3$ and PO allocation (Theorem~\ref{thm:efxthreepoter}). 

With absolute identical utilities, we gave Algorithm~\ref{alg:minimax} for computing an EFX and PO allocation (Theorem~\ref{thm:efxpoid}). With ternary utilities, this algorithm also returns such an allocation (Theorem~\ref{thm:efxpoter}). We further proved two impossibilities results (Propositions~\ref{pro:impefxthree} and~\ref{pro:impefxzero}). In particular, with ternary identical utilities, an EFX$_0$ allocation, or an EFX$^3$ allocation might not exist. We leave for future work two very interesting open questions with general utilities. Table~\ref{tab:results} contains our results. 

\begin{table*}[b]
\centering
\caption{Key: $\checkmark$-possible, $\times$-not possible, $\P$-polynomial time, $\alpha,\beta\in\mathbb{R}_{>0}:\alpha\neq\beta$.}
\label{tab:results}
\begin{tabular}{|c|C|C|C|C|C|C|C|C|C|C|C|C|}
\hline

\multirow{2}{*}{property} & \multicolumn{3}{C|}{general} & \multicolumn{3}{C|}{ident. \& abs.} & \multicolumn{3}{C|}{$-\alpha/0/\beta$} & \multicolumn{3}{C|}{$-\alpha/0/\alpha$} \\ 

 & \multicolumn{3}{C|}{utilities} & \multicolumn{3}{C|}{utilities} & \multicolumn{3}{C|}{utilities} & \multicolumn{3}{C|}{utilities} \\ \hline

EF1$^3$ & \multicolumn{3}{c}{$\checkmark$, $\P$ (Thm~\ref{thm:efonethreegen})} & \multicolumn{3}{c}{} & \multicolumn{3}{c}{}  & \multicolumn{3}{c|}{} \\ \cline{1-10}

EF1$^3$ \& PO & \multicolumn{3}{c|}{open} & \multicolumn{3}{c|}{$\checkmark$, $\P$ (Thm~\ref{thm:efonethreepoid})} & \multicolumn{3}{c|}{$\checkmark$, $\P$ (Thm~\ref{thm:efonethreepoter})} & \multicolumn{3}{c|}{}  \\ \cline{1-10}

EFX \& PO & \multicolumn{3}{c|}{open} & \multicolumn{3}{c|}{$\checkmark$, $\P$ (Thm~\ref{thm:efxpoid})} & \multicolumn{3}{c|}{$\checkmark$, $\P$ (Thm~\ref{thm:efxpoter})} & \multicolumn{3}{c|}{}  \\ \cline{1-10} 

EFX$^3$ & \multicolumn{3}{c}{} & \multicolumn{3}{c}{$\times$ (Prop~\ref{pro:impefxthree})} & \multicolumn{3}{c|}{} & \multicolumn{3}{c|}{}  \\ \cline{1-2}

EFX$^3$ \& PO & \multicolumn{3}{c}{} & \multicolumn{3}{c}{} & \multicolumn{3}{c|}{} & \multicolumn{3}{c|}{$\checkmark$, $\P$ (Thm~\ref{thm:efxthreepoter})}  \\ \hline

EFX$_0$ & \multicolumn{3}{c}{} & \multicolumn{3}{c}{$\times$ (Prop~\ref{pro:impefxzero})} & \multicolumn{3}{c}{} & \multicolumn{3}{c|}{}  \\ \hline

\end{tabular}
\end{table*}

\bibliographystyle{splncs04}
\bibliography{ki}

\begin{thebibliography}{10}
\providecommand{\url}[1]{\texttt{#1}}
\providecommand{\urlprefix}{URL }
\providecommand{\doi}[1]{https://doi.org/#1}

\bibitem{aleksandrov2015ijcai}
Aleksandrov, M., Aziz, H., Gaspers, S., Walsh, T.: Online fair division:
  analysing a food bank problem. In: Proceedings of the Twenty-Fourth {IJCAI}
  2015, Buenos Aires, Argentina, July 25-31, 2015. pp. 2540--2546 (2015)

\bibitem{amanatidis2020maximum}
Amanatidis, G., Birmpas, G., Filos-Ratsikas, A., Hollender, A., Voudouris,
  A.A.: Maximum {N}ash welfare and other stories about {EFX} (2020)

\bibitem{amanatidis2018}
Amanatidis, G., Birmpas, G., Markakis, V.: Comparing approximate relaxations of
  envy-freeness. In: Proceedings of the Twenty-Seventh International Joint
  Conference on Artificial Intelligence, {IJCAI} 2018, Stockholm, Sweden, July
  13-19, 2018. pp. 42--48 (2018)

\bibitem{aziz2019gc}
Aziz, H., Caragiannis, I., Igarashi, A., Walsh, T.: Fair allocation of
  indivisible goods and chores. In: Proceedings of the Twenty-Eighth
  International Joint Conference on Artificial Intelligence, {IJCAI-19}. pp.
  53--59. International Joint Conferences on Artificial Intelligence
  Organization (7 2019)

\bibitem{aziz2019popropone}
Aziz, H., Moulin, H., Sandomirskiy, F.: A polynomial-time algorithm for
  computing a {P}areto optimal and almost proportional allocation. CoRR
  \textbf{abs/1909.00740} (2019)

\bibitem{aziz2019group}
Aziz, H., Rey, S.: Almost group envy-free allocation of indivisible goods and
  chores. CoRR  \textbf{abs/1907.09279} (2019)

\bibitem{barman2018fe}
Barman, S., Krishnamurthy, S.K., Vaish, R.: Finding fair and efficient
  allocations. In: Proceedings of the 2018 ACM Conference on EC 2018. pp.
  557--574. ACM, New York, NY, USA (2018)

\bibitem{barman2018}
Barman, S., Krishnamurthy, S.K., Vaish, R.: Greedy algorithms for maximizing
  {N}ash social welfare. In: Proceedings of the 17th International Conference
  on Autonomous Agents and MultiAgent Systems, {AAMAS} 2018, Stockholm, Sweden,
  July 10-15, 2018. pp. 7--13 (2018)

\bibitem{bogomolnaia2019div}
Bogomolnaia, A., Moulin, H., Sandomirskiy, F., Yanovskaia, E.: Dividing bads
  under additive utilities. Social Choice and Welfare  \textbf{52}(3),
  395--417 (2019)

\bibitem{bogomolnaia2016gab}
Bogomolnaia, A., Moulin, H., Sandomirskiy, F., Yanovskaya, E.: Dividing goods
  and bads under additive utilities. CoRR  \textbf{abs/1610.03745} (2016)

\bibitem{bogomolnaia2017comp}
Bogomolnaia, A., Moulin, H., Sandomirskiy, F., Yanovskaya, E.: Competitive
  division of a mixed manna. Econometrica  \textbf{85}(6),  1847--1871 (2017)

\bibitem{brams1996}
Brams, S.J., Taylor, A.D.: Fair division -- from cake-cutting to dispute
  resolution. Cambridge University Press (1996)

\bibitem{budish2011}
Budish, E.: The combinatorial assignment problem: Approximate competitive
  equilibrium from equal incomes. Journal of Political Economy
  \textbf{119}(6),  1061--1103 (2011)

\bibitem{caragiannis2012}
Caragiannis, I., Kaklamanis, C., Kanellopoulos, P., Kyropoulou, M.: The
  efficiency of fair division. Theory of Computing Systems  \textbf{50}(4),
  589--610 (May 2012)

\bibitem{caragiannis2016}
Caragiannis, I., Kurokawa, D., Moulin, H., Procaccia, A.D., Shah, N., Wang, J.:
  The unreasonable fairness of maximum {N}ash welfare. In: Proceedings of {ACM}
  Conference on {EC} '16, Maastricht, The Netherlands, July 24-28, 2016. pp.
  305--322 (2016)

\bibitem{chevaleyre2006}
Chevaleyre, Y., Dunne, P., Endriss, U., Lang, J., Lemaitre, M., Maudet, N.,
  Padget, J., Phelps, S., Rodrguez-Aguilar, J., Sousa, P.: Issues in multiagent
  resource allocation. Informatica  \textbf{30},  3--31 (2006)

\bibitem{dobzinski2013}
Dobzinski, S., Vondr\'{a}k, J.: Communication complexity of combinatorial
  auctions with submodular valuations. In: Proceedings of the Twenty-fourth
  Annual ACM-SIAM Symposium on Discrete Algorithms. pp. 1205--1215. SODA '13,
  Society for Industrial and Applied Mathematics, Philadelphia, PA, USA (2013)

\bibitem{dubins1961}
Dubins, L.E., Spanier, E.H.: How to cut a cake fairly. The American
  Mathematical Monthly  \textbf{68}(1P1),  1--17 (1961)

\bibitem{foley1967}
Foley, D.K.: Resource allocation and the public sector. Yale Economic Essays
  \textbf{7}(1),  45--98 (1967)

\bibitem{steinhaus1948}
Hugo, S.: The problem of fair division. Econometrica  \textbf{16},  101--104
  (1948)

\bibitem{kyropoulou2019}
Kyropoulou, M., Suksompong, W., Voudouris, A.: Almost envy-freeness in group
  resource allocation. In: Proceedings of the Twenty-Eighth International Joint
  Conference on Artificial Intelligence, {IJCAI-19}. pp. 400--406.
  International Joint Conferences on Artificial Intelligence Organization (08
  2019)

\bibitem{lipton2004}
Lipton, R.J., Markakis, E., Mossel, E., Saberi, A.: On approximately fair
  allocations of indivisible goods. In: Proceedings of the 5th {ACM} Conference
  on {EC}, New York, NY, USA, May 17-20, 2004. pp. 125--131 (2004)

\bibitem{moulin2003}
Moulin, H.: Fair division and collective welfare. MIT Press (2003)

\bibitem{pareto1896}
Pareto, V.: Cours d'\'{E}conomie politique. Professeur \'{a} l'Universit\'{e}
  de Lausanne. Vol. I. Pp. 430. 1896. Vol. II. Pp. 426. 1897. Lausanne: F.
  Rouge  (1897)

\bibitem{plaut2018}
Plaut, B., Roughgarden, T.: Almost envy-freeness with general valuations. In:
  Proceedings of the 29th Annual {ACM-SIAM} Symposium on Discrete Algorithms,
  {SODA} 2018, New Orleans, LA, USA, January 7-10, 2018. pp. 2584--2603 (2018)

\bibitem{sandomirskiy2019minimal}
Sandomirskiy, F., Segal{-}Halevi, E.: Fair division with minimal sharing. CoRR
  \textbf{abs/1908.01669} (2019)

\bibitem{young1995}
Young, H.P.: Equity -- in theory and practice. Princeton University Press
  (1995)

\end{thebibliography}

\appendix

\section{A complete version of Algorithm~\ref{alg:mdrr}}\label{sec:alg}

For reasons of space, we presented a short version of Algorithm~\ref{alg:mdrr} in the main text. We present in here a complete version of it.

\setcounter{algorithm}{0}
\begin{algorithm}
\caption{An EF1$^3$ allocation.}
\begin{algorithmic}[1]
\Procedure{Modified Double Round-Robin}{$[n],[m],(u_1,\ldots,u_n)$} 
\State $M^0\gets\lbrace o\in [m]|\forall b\in [n]:u_b(t)\leq 0,\exists c\in [n]:u_c(t)=0\rbrace$
\State Allocate each item from $M^0$ to an agent who has utility $0$ for it. We let $A$ denote this allocation.
\State $M^-\gets\lbrace o\in [m]\setminus M^0|\forall a\in [n]: u_a(o)<0\rbrace$
\State Suppose $|M^-|=qn-p$ for some $q,p\in\mathbb{N}$ with $p<n$. Create $p$ ``fake'' dummy items for which each agent has utility $0$, and add them to $M^-$. Hence, $|M^-|=qn$.
\State Let the agents come in a round-robin sequence $(1,\ldots,n-1,n)$ and pick their most preferred item in $M^-$ until all items in it are allocated. 
\State $M^+\gets\lbrace o\in [m]\setminus M^0|\exists a\in [n]: u_a(o)>0\rbrace$
\State Let the agents come in a round-robin sequence $(n,n-1,\ldots,1)$ and pick their most preferred item in $M^+$ until all items in it are allocated. If an agent has no available item which gives them strictly positive utility, they pretend to pick a ``fake'' dummy item for which they have utility $0$.
\State Remove the ``fake'' dummy items from the current allocation and return the resulting allocation. We let $B$ denote this allocation.
\State \Return $(A_1\cup B_1,\ldots,A_n\cup B_n)$ 
\EndProcedure
\end{algorithmic}
\end{algorithm}

\end{document}